\documentclass[fp]{jpsj3}

\usepackage{txfonts}

\usepackage{textgreek} %
\usepackage{siunitx} %
\usepackage{microtype}

\newcommand{\irsn}{\mbox{\textbeta-IrSn\textsubscript{4}}}

\title{%
    Linear Magnetoresistance and Type-I Superconductivity in \irsn%
}

\author{%
    Nazir~Ahmad$^1$, Shunsuke~Shimada$^1$, Takumi~Hasegawa$^2$, Hiroto Suzuki$^1$,  Md~Asif~Afzal$^1$, Naoki~Nakamura$^1$, Ryuji~Higashinaka$^1$, Tatsuma~D.~Matsuda$^1$, and Yuji~Aoki$^1$\thanks{corresponding author: aoki@tmu.ac.jp}
}

\inst{%
    $^1$Department of Physics, Tokyo Metropolitan University, Hachioji, Tokyo 192--0397, Japan\\
    $^2$Graduate School of Advanced Science and Engineering, Hiroshima University, Higashi-Hiroshima 739--8521, Japan %
}

\abst{ %
    Layered material \irsn\ ($I4_1/acd$, $D^{20}_{4h}$, \#142), whose electron bands have symmetry-enforced Dirac points, was investigated using high-quality single crystals.
    It exhibits a pronounced linear field-dependence of magnetoresistance (LMR), which cannot be explained by currently existing models.
    Structures in the field-angle dependence of magnetoresistance and Hall resistivity are attributable to the Fermi surface topology; the presence of open orbits is inferred.
    At the superconducting (SC) transition, the specific-heat jump exhibits a significant increase in applied fields, revealing the type-I SC nature.
    This feature is attributable to the high Fermi velocity of linearly dispersive multibands.
    To clarify the mechanism of the puzzling LMR, investigations into the topological nature of those multibands in applied fields are highly desired. %
}

\begin{document}
    \maketitle
    \section{Introduction}
    Materials containing band touching points, known as Dirac or Weyl points, described by the massless Dirac equation have attracted considerable attention in this decade because of their exotic electronic properties stemming from the topologically nontrivial features of the bands~\cite{yang2014,armitage2018,gorbar2021,lv2021}.
    These include the formation of Fermi-arc surface states~\cite{kargarian2016,clark2018,breitkreiz2019}, chiral anomalies, e.g., negative longitudinal magnetoresistance (MR)~\cite{xiong2015,jia2016,armitage2018,lv2021,ong2021}, a non-trivial Berry phase observed in quantum oscillations~\cite{xiao2010,he2014,fei2017,vanderbilt2018,guo2021,alexandradinata2023}, largely enhanced orbital diamagnetism~\cite{koshino2010,suetsugu2021},  high carrier mobility caused by suppressed backscattering of conducting electrons,~\cite{liang2014} and more.
    
    Linear magnetoresistance (LMR), a phenomenon where resistivity increases linearly with magnetic field $H$ over a broad temperature and field range, is an unusual magnetotransport property often observed in Dirac semimetals, e.g., Cd$_3$As$_2$~\cite{liang2014,feng2015}, TlBiSSe~\cite{novak2015}, and Weyl semimetals, e.g., NbP~\cite{niemann2017}, TaP~\cite{leahy2018}.
    To account for the phenomena, many theoretical ideas have been proposed, such as the ``quantum limit'' of a Dirac electron state (only the lowest Landau level being partially occupied in high enough fields)~\cite{abrikosov1998,abrikosov2000}, $H$-linear slowing down of carriers in bands that have Berry curvature~\cite{zhang2022}, and strongly anisotropic (momentum-dependent) scattering causing impeded orbital motion in correlated electron systems~\cite{hinlopen2022}.
    However, the mechanisms are still debated and require further clarification.

\begin{figure}
        \centering
        \includegraphics[width=.9\linewidth]{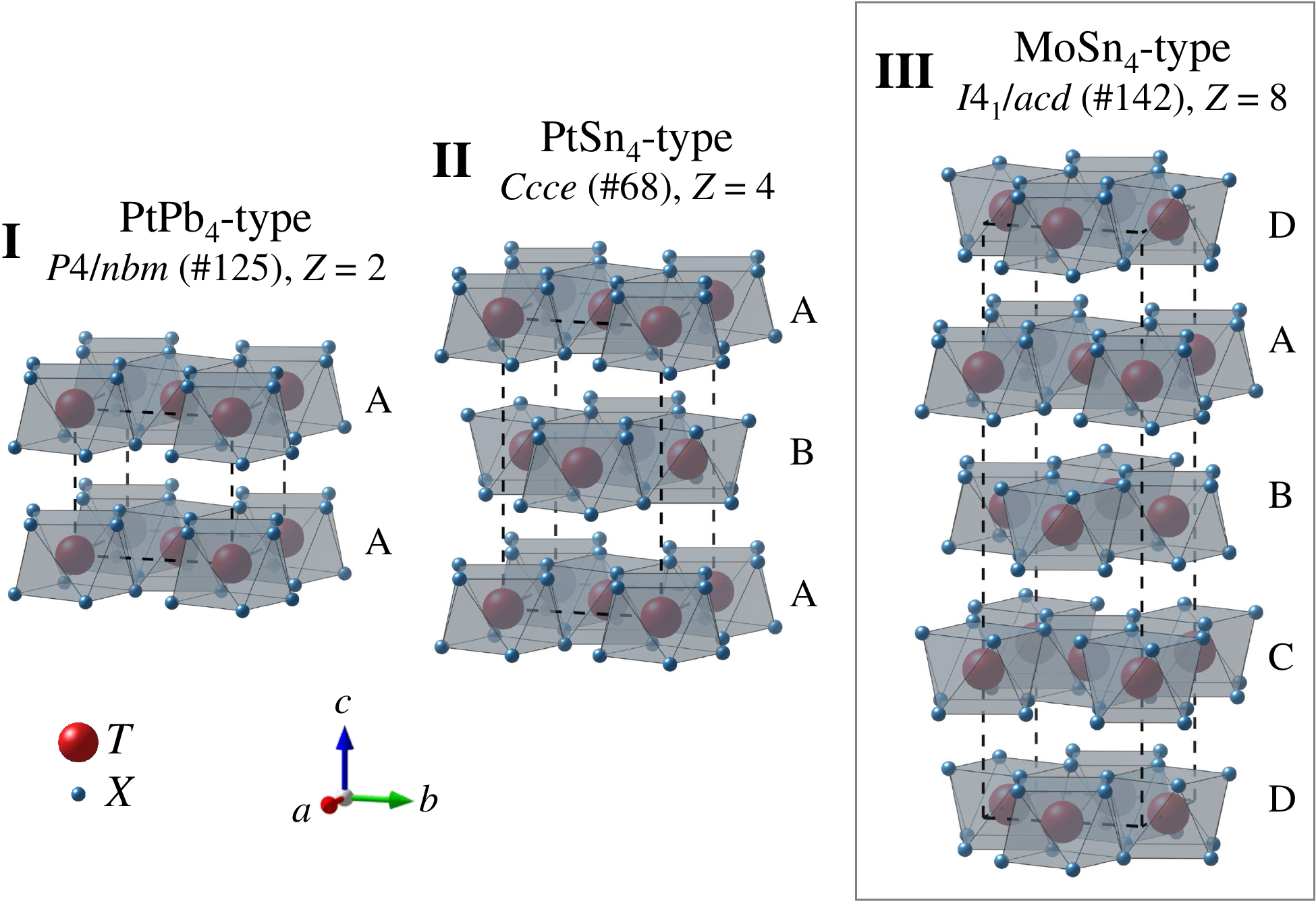}
        \caption{(Color online) Family of \textit{TX}\textsubscript{4} structures derived from the CuAl$_2$ (C16) structure.
            The $T$ atoms are coordinated by eight $X$ atoms forming a square antiprism.
            \textit{TX}\textsubscript{8} square antiprisms form a layer by sharing common edges.
            The crystal structures of three members (I-III) show a different sequence of PtPb$_4$-type layer stacking along the tetragonal [001] direction (the \textit{c}-axis), viz., {A} (I), {AB} (II), and {ABCD} (III)~\cite{nordmark2002}.
            \irsn\ belongs to III, commonly known as MoSn$_4$-type (body-centered tetragonal $I4_1/acd$, $D^{20}_{4h}$, \#142).
        }
        \label{fig:crystal}
    \end{figure}

    To explore these novel Dirac-electron phenomena, several members of the \textit{TX}\textsubscript{4} family ($T$: transition metal, $X$: p-block element), featuring a layer-stacking crystal structure, have presented a unique experimental platform.
    Figure~\ref{fig:crystal} illustrates the crystal structures of three such members, which have different sequences of layer stacking along the [001] direction, namely {A} (I: PtPb$_4$-type), {AB} (II: PtSn$_4$-type), and {ABCD} (III: MoSn$_4$-type)~\cite{nordmark2002}. 
    Some compounds in this family have been reported to possess Dirac electron states: (I) Dirac nodal-line with significant Rashba splitting in PtPb$_4$~\cite{lee2021,wu2022} and (II) Dirac node arc in PtSn$_4$~\cite{wu2016}.
    However, the topological features of materials in III, including \irsn\ \cite{nordmark2002,tran2013,vergniory2019,mai2022}, remain largely unexplored.
    \irsn\ was reported to be a superconductor with a superconducting (SC) transition temperature of $T_\text{C}=0.9$~K, although the nature of SC has not been well clarified~\cite{tran2013}.
    
    In this {paper}, we report the synthesis of high-quality single crystals of \irsn.
    Measurements of the electrical resistivity provide the first observation of LMR in this material.
    In addition, specific heat measurements reveal largely enhanced sharp jumps at the SC transition in applied fields, providing unambiguous evidence for the type-I nature of the SC state.
    Possible reasons for these findings are discussed together with the electronic band structure obtained from first-principles calculations.
    
    \section{Experimental Details}
    Single crystals of \irsn\ were grown using the self-flux method.
    A mixture of iridium powders (4N, 99.99\%) and tin grains (5N, 99.999\%) was sealed in an evacuated quartz tube with an atomic ratio of 1:20.
    This mixture was heated to 950~$^\circ$C and kept at that temperature for a day. Next, it was gradually cooled to 700~$^\circ$C at a rate of 1.7~$^\circ$C/h followed by a subsequent rapid quenching down to room temperature to prevent the formation of \textalpha-IrSn\textsubscript{4}, which is stable at low temperatures.~\cite{nordmark2002, nakamura2023}
    Excess tin was removed by etching in diluted hydrochloric acid (HCl).
    
    Single crystal X-ray diffraction analysis was performed using a Rigaku XtaLABmini with graphite monochromated Mo-$K\alpha$ radiation. 
    The crystals were confirmed to have the MoSn$_4$-type crystal structure ($I4_1/acd$, \#142).
    See Supplemental Material (SM) for details~\cite{supplement}.
    
    The electrical resistivity ($\rho$), Hall resistivity ($\rho_\text{H}$) and specific heat ($C$) were measured using a Quantum Design (QD) Physical Property Measurement System (PPMS) equipped with a Helium-3 cryostat and a rotator option.
    The sample used for the specific heat measurement had dimensions of {1.50~mm $\times$ 1.50~mm $\times$ 0.23~mm}. The [001] direction is normal to the largest plane, resulting in a demagnetizing factor of $N=0.81 (0.15)$ for $H\parallel [001]$ ($H\parallel [110]$).~\cite{prozorov2018}
    
    \section{Electronic Band Structure}
    The first principles calculation of \irsn\ was performed using the ABINIT package~\cite{gonze2020,romero2020} with local density approximation (LDA), implemented in the projector augmented wave~\cite{bloechl1994,torrent2008} (PAW) framework. 
    The exchange--correlation potential by Perdew and Wang~\cite{perdew1996} and the spin--orbit coupling were incorporated; refer to SM for further details~\cite{supplement}.
    
\begin{figure}
        \centering
        \includegraphics[width=.9\linewidth]{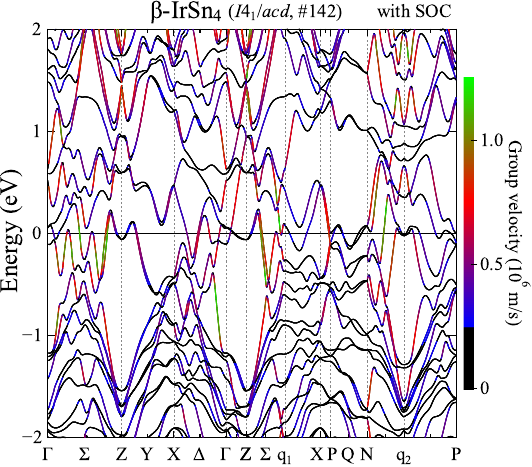}
        \caption{(Color online) The electronic energy band structure of \irsn with spin--orbit coupling (SOC). 
            The color scale indicates the group velocity. 
            At the Fermi energy $E_\text{F}$, numerous linear bands exhibiting notably high Fermi velocity $v_\text{F}$ can be observed, comparable to the velocities found in graphene ($1\times10^6$~m/s). For details see SM~\cite{supplement}.
        }
        \label{fig:dos}
    \end{figure}

    Recently, it has been clarified theoretically for tetragonal systems 
    that the symmetry of the crystal guarantees the existence of some Dirac points located in the Brillouin zone, which are called {\it symmetry-enforced Dirac points}~\cite{hirschmann2021}.
    For $I4_1/acd$ (\#142) with the lattice parameters $c > a$, the Dirac points can occur at X, Z, N, and P on the Brillouin zone boundary and accidental Dirac points on the lines of X--P, X--Y, N--P, and $\Gamma$--Z~\cite{hirschmann2021}.
    In fact, corresponding Dirac points appear in the calculated band structure shown in Fig.~\ref{fig:dos}.
    At P point, one typical example exists in close proximity to the Fermi energy $E_\text{F}$ (at $-$0.1~eV), forming a small electron pocket (see SM for details~\cite{supplement}).
    On the other hand, \irsn\ is a compensated metal, possessing an equal number of electrons and holes.
    However, it is difficult to determine whether \irsn\ belongs to the category of {\it Dirac semimetals} because of the presence of other $E_\text{F}$-crossing multibands that appear to lack Dirac points.
    Nevertheless, those bands have a remarkable feature.
    Many of them exhibit almost linear and steep dispersions; the Fermi velocity $v_\text{F}$  reaches $1.2\times10^6$~m/s, which is comparable to $1\times10^6$~m/s of graphene~\cite{castroneto2009}.
    
    \section{Results and Discussion}
    \begin{figure}
        \centering
        \includegraphics[width=.7\linewidth]{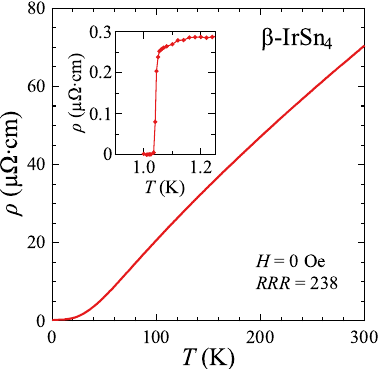}
        \caption{(Color online) The zero-field temperature dependence of resistivity $\rho(T)$ of \irsn. 
            The residual resistivity ratio $RRR \equiv \rho(300~\text{K})/\rho(1.2~\text{K})=238$ is the highest in the literature. 
            Inset: a steep drop at 1.05~K confirms the superconducting (SC) transition.}
        \label{fig:rho}
    \end{figure}
    
    \subsection{Magnetoresistance and Hall resistivity}
    Figure~\ref{fig:rho} displays the temperature dependence of resistivity $\rho(T)$, which exhibits a metallic behavior over the entire temperature range.
    The residual resistivity ratio $RRR=\rho(300~\text{K})/\rho(1.2~\text{K})$ is ${\sim}240$, the highest in the literature, attesting to the high quality of the crystals.
    A sharp drop at around 1.05~K (slightly higher than 0.9~K, reported in the earlier work~\cite{tran2013}) confirms the SC transition.
    
\begin{figure*}[h]
        \centering
        \includegraphics[width=.9\linewidth]{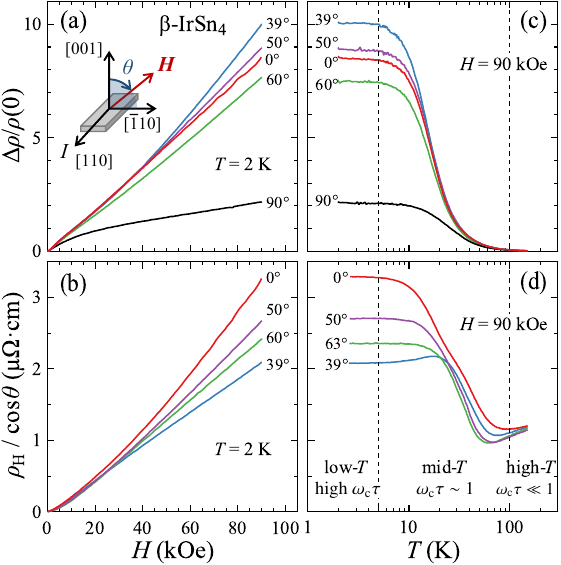}
        \caption{(Color online) (a,b) $H$- and (c,d) $T$-dependence of transverse magnetoresistance (MR) $\Delta \rho/\rho(0)$ (see text for the definition) and Hall resistivity $\rho_\text{H}/\cos \theta$ in \irsn\ for different magnetic field orientations with the current $I \parallel [110]$.
            The magnetic field is rotated in the (110) plane as illustrated in the inset.
            MR showcases a non-saturating quasi-linear $H$-dependence. %
            $T$-dependence of MR and $\rho_\text{H}/\cos \theta$ implies that the observed data can be separated into three temperature regions (for details see text).
        }
        \label{fig:MR}
    \end{figure*}

    Figure~\ref{fig:MR}(a,c) show $H$- and $T$-dependence of transverse  MR defined by $\Delta \rho / \rho(0) \equiv [\rho(H)-\rho(0)]/\rho(0)$ with $\rho(0)\equiv\rho(H=0)$ for different magnetic field directions with the current $I \parallel [110]$.
    At 2~K, MR exhibits almost a linear-in-$H$ behavior in the measured field range reaching $\Delta \rho / \rho(0) \sim 10$ in 90~kOe without any signs of saturation. The enhancement of MR develops below ${\sim}100$~K as displayed in Fig.~\ref{fig:MR}(c).
    
    The LMR is quite surprising in \irsn\ as it is a compensated metal and is therefore expected to exhibit $H^2$ dependence according to the conventional electron-transport theories~\cite{pippard1989, chambers1990}.
    Several scenarios have been proposed to account for LMR.
    (i) In inhomogeneous materials, {\it mobility fluctuations} can lead to LMR as proposed for black phosphorus~\cite{hou2016}.
    This scenario is not applicable to the high-quality \irsn\ single crystals with  high $RRR$ ($\sim$240).
    (ii) {\it the quantum limit of a Dirac electron state} (the lowest Landau level being filled in high enough fields)~\cite{abrikosov1998,abrikosov2000} is ruled out because LMR appears down to $H \sim 1$ kOe~\cite{quantumlimit}.
    (iii) {\it Strongly anisotropic (momentum-dependent) scattering} causing impeded orbital motion is unlikely since \irsn\ is not a correlated electron system with hot spots on the Fermi surface~\cite{hinlopen2022}.
    (iv) A theory suggests that, if bands have Berry curvature, carriers can experience $H$-linear slowing down, resulting in LMR~\cite{zhang2022}.
    As shown in Fig.~\ref{fig:dos}, \irsn\ has linearly dispersive multiple bands around $E_\text{F}$.
    Some of these bands are extrapolated to a symmetry-enforced Dirac point although they are intersected by other bands in between.
    If non-zero Berry curvature appears in applied fields by the splitting of Dirac points into Weyl points (breaking time-reversal symmetry) as discussed for Cd$_3$As$_2$~\cite{jia2016}, which belongs to the same space group, this mechanism can explain the LMR.
    This conjecture requires theoretical verification in the future.
    
    Hall resistivity $\rho_\text{H}(H)$ (see Fig.~\ref{fig:MR}(b)) shows an anomalous non-linear behavior and does not follow the $\cos \theta$ dependence; the data for $\theta=0^{\circ} \sim 60^{\circ}$ do not fall into a single curve.
    $\rho_\text{H}>0$ suggests that holes have higher mobility than electrons.
    The temperature dependence of MR and $\rho_\text{H}$ (see Fig.~\ref{fig:MR}(c,d)) indicates that the temperature range can be separated into three regions.
    Below $\sim$5~K (low-$T$ region), both MR and $\rho_\text{H}$ are saturated (high $\omega_\text{c} \tau$, where $\omega_\text{c}$ is the cyclotron frequency and $\tau$ the relaxation time).
    Above $\sim$100~K (high-$T$ region), MR is suppressed and $\rho_\text{H}/\cos \theta$ is saturated to $\sim$ 1.2~\textmu$\Omega\cdot$cm, which greatly differs from that observed in the low-$T$ region, indicating that $\rho_\text{H}$ is determined by the local curvature of the Fermi surface in the \textbf{\textit{k}} space ($\omega_\text{c} \tau \ll 1$).
    In the mid-$T$ region, both MR and $\rho_\text{H}$ change with temperature through the $T$ dependence of $\tau$.
    
\begin{figure}
        \centering
        \includegraphics[width=.65\linewidth]{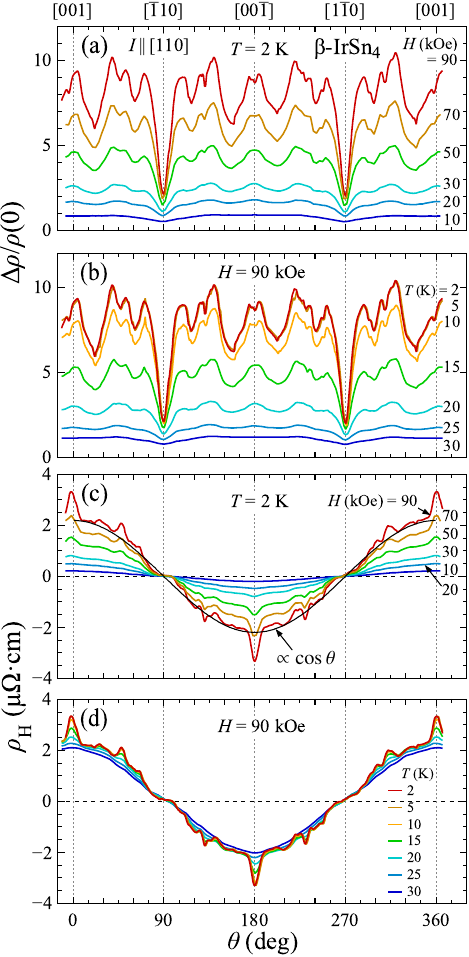} %
        \caption{(Color online) (a,b) MR $\Delta \rho/\rho(0)$ and (c,d) $\rho_\text{H}$ as a function of field-angle $\theta$ for different $H$ at 2~K and different $T$ in 90~kOe.
            Refer to the inset of Fig.~\ref{fig:MR} for the field configuration.
            A complex structure develops in both quantities at low $T$ and high $H$, probably reflecting the Fermi surface topology of \irsn.}
        \label{fig:angMR}
    \end{figure}

    The field-angle dependence of MR and $\rho_\text{H}$ are shown in Fig.~\ref{fig:angMR}: (a,c) for different $H$ at 2~K and (b,d) for different $T$ in 90~kOe.
    These figures demonstrate a clear development of $\theta$-dependent structures with increasing $H$ and decreasing $T$, which most likely reflects the Fermi surface topology.
    The most prominent feature of MR is the sharp dip at $\theta=90^{\circ}$ and $270^{\circ}$.
    This observation suggests the existence of open orbits along the [001] direction (perpendicular to both $I \parallel [110]$ and $H \parallel [\bar110]$~\cite{chambers1990}), which should be confirmed by de~Haas--van~Alphen (dHvA) or angle-resolved photoemission spectroscopy (ARPES) measurements.
    
    \subsection{Specific heat}
\begin{figure*}[h]
        \centering
        \includegraphics[width=.9\linewidth]{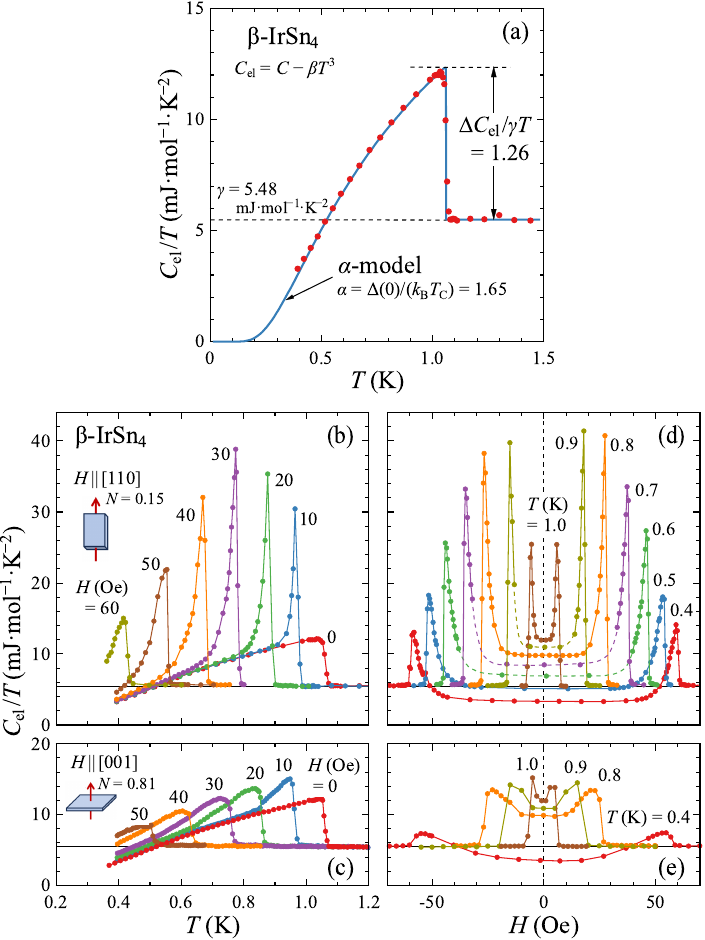} %
        \caption{(Color online) (a) Temperature dependence of the electronic specific heat $C_\text{el}$ of \irsn\ in zero field. 
            The best fitting $\alpha$-model curve with $\alpha=1.65$ is shown by the blue line.
            $T$- and $H$-dependence of $C_\text{el}$ in applied magnetic fields (b,d) for $H \parallel [110]$ and (c,e) for $H \parallel [001]$.
            A remarkable increase of the specific heat jump $\Delta C_\text{el}$ is observed in applied fields, especially for $H\parallel [110]$ around $H=20$~Oe--$30$~Oe. 
            Dashed lines are a guide to the eye.
        }
        \label{fig:HC}
    \end{figure*}

    Specific heat $C(T)$  in the normal state yields the Sommerfeld coefficient $\gamma=5.48$~mJ/(mol$\cdot$K$^2$) and the Debye temperature $\Theta_\text{D}=256$~K (see SM for details~\cite{supplement}).
    Figure~\ref{fig:HC}(a) depicts the temperature dependence of the electronic specific heat $C_\text{el}/T = C - \beta T^3$ in zero field.
    At $T_\text{C}=1.06$~K, a sharp jump is observed and the size of the jump $\Delta C_\text{el}/\gamma T_\text{C}=1.26$ is slightly smaller than the expected 1.43 from the BCS theory. This indicates a weak-coupling nature of SC in \irsn.
    The $C_\text{el}(T)$ data below $T_\text{C}$ can be well fitted by a phenomenological $\alpha$ model with $\alpha \equiv \Delta(0)/(k_\text{B}T_\text{C})=1.65$, where $\Delta(0)$ is the SC energy gap at $T=0$ and $k_\text{B}$ the Boltzmann constant~\cite{padamsee1973, johnston2013}.
    The $\alpha$ value, which is smaller than 1.764 expected from the BCS theory, may be attributed to anisotropy in the SC gap~\cite{johnston2013}.
    
    Figures~\ref{fig:HC}(b-e) illustrate the $T$- and $H$-dependence of $C_\text{el}/T$ for $H\parallel [110]$ and $H\parallel [001]$.
    The transition temperature indicated by the $C_\text{el}$ jump shifts to lower temperatures with increasing field.
    The most notable feature in these figures is largely enhanced $\Delta C_\text{el}$ at the SC transition in applied fields, especially for $H \parallel [110]$. Around $H=20$~Oe--$30$~Oe, $\Delta C_\text{el}/\gamma T$ reaches 6.5.
    This enhancement of $\Delta C_\text{el}$ provides unambiguous evidence for the occurrence of type-I SC in \irsn\ (see the discussion below).
    Similar enhancement was reported for type-I superconductors, e.g., PdTe$_2$ ($T_\text{C}=1.3$~K)~\cite{salis2021}, ScGa$_3$ (2.1~K)~\cite{svanidze2012}, LuGa$_3$ (2.2~K)~\cite{svanidze2012}, and YbSb$_2$ (1.3~K)~\cite{zhao2012}, although $\Delta C_\text{el}/\gamma T$ is much smaller than those observed in \irsn.
    
\begin{figure}
        \centering
        \includegraphics[width=.8\linewidth]{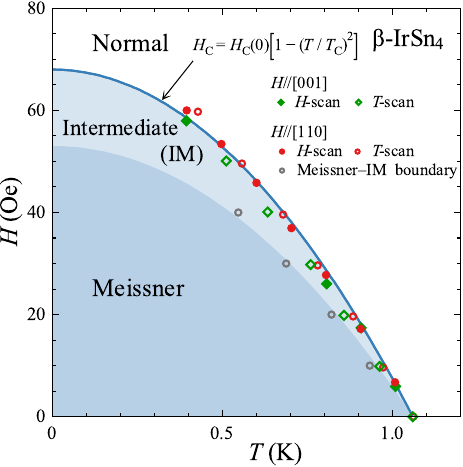}
        \caption{(Color online) $H{-}T$ phase diagram of \irsn\ determined from the specific heat data shown in Fig.~\ref{fig:HC}.
            The critical field $H_\text{C}(T)$, as observed, is almost isotropic and agrees well with the empirical formula for type-I SC.
            The intermediate (IM) phase, determined from specific heat measurement for $H\parallel [110]$, is presented in a lightly  shaded blue region.
        }
        \label{fig:HT}
    \end{figure}

    Figure~\ref{fig:HT} displays the $H{-}T$ phase diagram of \irsn, which is generated from the data shown in Fig.~\ref{fig:HC}.
    The critical field $H_\text{C}(T)$ is almost the same for $H \parallel [110]$ and $H \parallel [001]$.
    The isotropic feature of $H_\text{C}(T)$ is consistent with the type-I SC nature. However, if the material under investigation were type-II SC, the anisotropic coherence length, expected for layered materials, would lead to anisotropic upper critical field $H_\text{C2}(T)$.
    In fact, the experimental $H_\text{C}(T)$ agrees well with the empirical formula for type-I SC: $H_\text{C} (T) = H_\text{C}(0) \left[1 - (T/T_\text{C})^2\right]$ with $H_\text{C}(0) = 68$~Oe and $T_\text{C} = 1.06$~K, as depicted in Fig.~\ref{fig:HT}~\cite{trandata}.
    
    The $\Delta C_\text{el}$ enhancement in applied fields can be attributed to the partial superconducting--normal (SN) state transition (or change in the normal-state volume fraction) occurring in the intermediate (IM) phase.
    The boundary between Meissner and IM phases is estimated by a simple model: $H_\text{MI}(T) = H_\text{C} (T)(1 - N)$ with $N$ being the demagnetizing factor.
    For $H\parallel [110]$ $(N=0.15)$, $H_\text{MI}(T)$ is shown in Fig.~\ref{fig:HT}.
    Experimentally, $H_\text{MI}(T)$ is obtained from the data shown in Figs.~\ref{fig:HC}(b-e).
    Tentatively, $[C_\text{el}(T_\text{MI}, H_\text{MI})-C_\text{el}(T_\text{MI},0)]/\Delta C_\text{el}(T_\text{C}(H_\text{MI}), H_\text{MI})=0.05$ is used and plotted in Fig.~\ref{fig:HT} for $H\parallel [110]$.
    The data points are reasonably consistent with the abovementioned model curve.
    The latent heat associated with the type-I SC transition in applied fields is continuously distributed in the IM phase.
    Therefore, the smaller IM phase region leads to higher $\Delta C_\text{el}$ for $H\parallel [110]$.
    
    The type-I nature of SC is attributable to high $v_\text{F}$.
    With $v_\text{F} = 1 \times 10^6$~m/s, the Pippard coherence length $\xi_0 =0.18 \hbar v_\text{F}/(k_\text{B} T_\text{C}) = 1.3$~\textmu m is evaluated~\cite{tinkham1996}.
    Using $\Phi_0=(2/3)^{1/2}\pi^2 \xi_0 \lambda_\text{L}(0)H_\text{C}(0)$, where $\Phi_0$ is the magnetic flux quantum and $\lambda_\text{L}(0)$ the penetration depth at $T=0$, Ginzburg--Landau parameter $\kappa =0.96 \lambda_\text{L}(0)/\xi_0 = 0.02  < \frac{1}{\sqrt{2}}$ is estimated in the pure limit~\cite{tinkham1996}.
    
    Most of the known non-elemental superconductors belong to type-II.
    Only a few are reported to be type-I: for example, in addition to the abovementioned ones, Al$_6$Re ($T_\text{C}=0.7$~K)~\cite{peets2019}, BeAu (3.2~K)~\cite{singh2019c}, Rh$_2$Ga$_9$ (2~K)~\cite{shibayama2007}, and LiPd$_2$Ge (1.96~K)~\cite{gornicka2020} and Ag$_5$Pb$_2$O$_6$ (0.05~K)~\cite{yonezawa2005}.
    Note that, among Dirac semimetals, only PdTe$_2$ falls under this category~\cite{leng2017,salis2021}.
    
    \section{Summary}
    We investigated the electronic transport and thermodynamic properties of layered material \irsn\ using high-quality single crystals of $RRR \simeq 240$.
    We revealed a pronounced  LMR.
    Although the electron bands of \irsn\ have symmetry-enforced Dirac points, LMR cannot be explained by the quantum limit scenario of the Dirac points, as the quantum limit cannot be reached in the measured field region.
    In the field-angle dependence of both MR and Hall resistivity, anomalous structures appear, probably reflecting the Fermi surface topology.
    The large enhancement of specific heat jump at the superconducting (SC) boundary provides compelling evidence for the type-I SC nature, which is probably associated with the high $v_\text{F}$ of the linearly dispersive multiple bands.
    One possible explanation for the puzzling LMR is the appearance of non-zero Berry curvature by the field-induced splitting of Dirac points into Weyl points, causing an $H$-linear slowing down of carriers~\cite{zhang2022}.
    It is highly desirable to clarify the topological nature of those bands around $E_\text{F}$ to test this hypothesis.
    LMR has been reported in many types of materials, among which multiband systems are rare.
    \irsn\ may provide a promising example of {multiband electron systems of a topologically nontrivial nature}, which has not been intensively studied.
    
    \begin{acknowledgments}
        The authors thank Prof.~Y.~\=Onuki and {Dr.~M.~M.~Hirschmann} for helpful discussions. This work was supported by JSPS KAKENHI Grant Numbers JP19H01839, JP22K03517, JP22H03522, JP23H04870, JP23K03332 and by Tokyo Metropolitan Government Advanced Research (Grant Number H31-1).  
    \end{acknowledgments}

\end{document}


\maketitle
    \clearpage
    \section{Single-Crystal X-Ray Diffraction Analysis}
    Single-crystal X-ray diffraction analysis was performed using a Rigaku XtaLABmini with graphite monochromated Mo-K\textalpha\ radiation. 
    The approximate size of single crystals was 0.1~mm~$\times$ 0.1~mm~$\times$~0.01~mm.
    The structural parameters refined using the program SHELXL\cite{robinson1988,sheldrick2007} are shown in Table~\ref{tab:structure}. 
    The atomic positions agree well with the previous results~\cite{nordmark2002, tran2013}.
    \begin{table}[h]
        \centering
        \caption{Crystallographic parameters of \irsn\ (space group: $I4_1/acd$,  \#142; origin choice 2) at room temperature determined by single-crystal X-ray measurements ($2\theta_\text{max}=54.9^\circ$; reflections collected: 3912; unique reflections: 266).
            $B_\text{eq}$ is the equivalent isotropic atomic displacement parameter.
            $R$ and $wR$ are reliability factors.
            The lattice parameters were evaluated from high-angle $(00l)$ and $(h\,h\,2h)$ reflections ($2\theta \approx 140^\circ$) using a Rigaku SmartLab diffractometer.
            Standard deviations in the positions of the least significant digits are given in parentheses.\\}
        \label{tab:structure}
        \begin{tabular}{@{}cccccc@{}}
            \toprule
            \multicolumn{2}{c}{\multirow{1}{*}{$I4_1/acd$ (origin choice 2)}} & \multicolumn{4}{c}{$a=b=6.3100(7)~\AA, c=22.7679(4)~\AA; V=906.53~\AA^3$}\\
            \multicolumn{2}{c}{$D^{20}_{4h},$ \#142, $Z = 8$}& \multicolumn{3}{c}{Position} & \multicolumn{1}{c}{\multirow{2}{*}{$B_\text{eq}$ (\AA$^2$)}} \\ \cmidrule{3-5}
            Atom & site & $x$ & $y$ & $z$ & \\ \midrule
            Ir & 8$b$ & 0 & 1/4 & 1/8 & 0.65(4) \\
            Sn & 32$g$ & 0.32657(14) & 0.07689(14) & 0.06142(4) & 0.81(4)\\\bottomrule
            \multicolumn{6}{l}{$R_1=5.2~\%,R_\text{all}=6.9~\%,wR_2=10.6~\%$}
        \end{tabular}
    \end{table}
    
    \clearpage
    \section{Band Structure Calculations}
    The first principles calculation of \irsn\ was performed using ABINIT package~\cite{gonze2020,romero2020} with local density approximation (LDA), and with projector augmented wave (PAW) method~\cite{bloechl1994,torrent2008}.
    The exchange-correlation potential prescribed by Perdew and Wang~\cite{perdew1996} was employed.
    The spin-orbit interaction was also taken into consideration.
    The partial waves and projector functions were generated by ATOMPAW~\cite{holzwarth2001} with valence-electron configurations of 6s$^1$5d$^8$ for Ir and 5s$^2$5p$^2$ for Sn. 
    The plane wave cutoff energy was 435 eV, and a $6\times6\times4$ Monkhorst--Pack \linebreak \textbf{\textit{k}}-point mesh of the conventional tetragonal Bravais lattice was applied.
    For metallic occupation, Fermi--Dirac smearing scheme was adapted with $T = 27.2$~meV. 
    The crystal structure was optimized prior to the calculation, resulting in optimized lattice parameters of $a=6.245$~\AA\ and $c = 22.619$~\AA, and an atomic position for Sn of (0.32730, 0.07687, 0.06159).
    
\begin{figure}
        \centering
        \includegraphics[width=.95\linewidth]{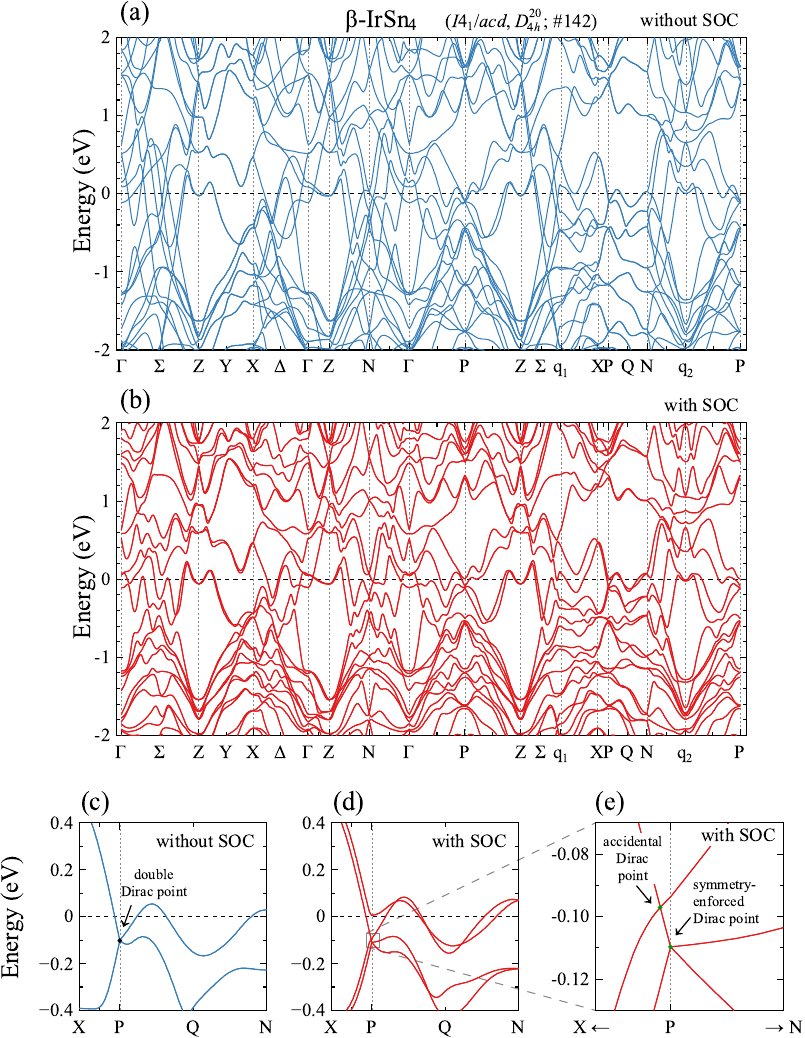}
        \caption{Electronic energy band structure of \irsn\ (a) without  spin-orbit coupling (SOC) and (b) with SOC. 
            The coordinates of the high symmetry points in the BZ are as follows: $\Gamma$(0,0,0)--Z(1,0,0)--X(0.5,0.5,0)--$\Gamma$(0,0,0)--Z(0,0,1)--N(0,0.5,0.5)--$\Gamma$(0,0,0)--P(0.5,0.5,0.5)--Z(1,1,1)--q$_1$(0.5,1,1)--X(0.5,0.5,1)--P(0.5,0.5,0.5)--N(0,0.5,0.5)--q$_2$(0,0,0.5)--P(0.5,0.5,0.5).
            The high symmetry points in the Brillouin zone can be found elsewhere~\cite{bradley2010}.  The band structure around the high-symmetry P point (c) without SOC and (d) with SOC. (e) In an expanded view (with SOC), a symmetry-enforced Dirac point ($E=-0.11$~eV) at P and an accidental Dirac point ($E=-0.097$~eV) near P on the X--P line are marked by solid green circles.}
        \label{fig:DFT}
    \end{figure}

    The calculated energy band structures of \irsn\ are shown in Fig.~\ref{fig:DFT}(a) without spin-orbit coupling (SOC) and in Fig.~\ref{fig:DFT}(b) with SOC.
    Additional data connecting high symmetry points, which are omitted in Fig.~2(a) of the main text, are added.
    Note that all the bands are at least doubly degenerate due to the time-reversal and inversion symmetry of the crystal.
    
    Many linearly dispersive bands appear near $E_\text{F}$, especially the regions of $\Gamma$--$\Sigma$--Z, X--$\Delta$--$\Gamma$, and Z--N--$\Gamma$ (this feature is clearer in the case of band structure without SOC). These bands have the Fermi velocity $v_\text{F}$ up to $1.2 \times 10^6$~m/s, which is comparable to $1\times10^6$~m/s of graphene~\cite{castroneto2009}.
    This feature may be attributed to the existence of many symmetry-enforced Dirac points on the boundary of the Brillouin zone, along with the four-times band folding along the $k_z$ direction, owing to the four-layer stacking along the [001] direction of the crystal structure.
    
    Figure~\ref{fig:DFT}(d) shows the electronic energy band structure around the high-symmetry P point with SOC; see Fig.~\ref{fig:DFT}(e) for an expanded view.
    At P point, a symmetry-enforced Dirac point appears at $E=-$0.11~eV.
    In addition, on the X--P line near P, there exists an accidental Dirac point at $E=-0.097$~eV, which is closer to the Fermi level $E_\text{F}$.
    The existence of these two Dirac points is in agreement with the theoretical study~\cite{hirschmann2021}.
    The band connected to the Dirac points forms a small electron pocket at P.
    Note that, however, the electrons of this pocket seem to have a minor contribution to the transport properties since $\rho_\text{H}>0$ (see Fig.4 in the main text) suggests that holes have higher mobility than electrons.
    In the absence of SOC, the band splitting disappears, and the two separate Dirac points merge into a {\it double Dirac point} (quadruply degenerate), around which the bands exhibit linear dispersion in all three directions, as presented in Fig.~\ref{fig:DFT}(c).

    \clearpage
    \section{Linear magnetoresistance (LMR): sample dependence}
\begin{figure}[h]
        \centering
        \includegraphics[width=.95\linewidth]{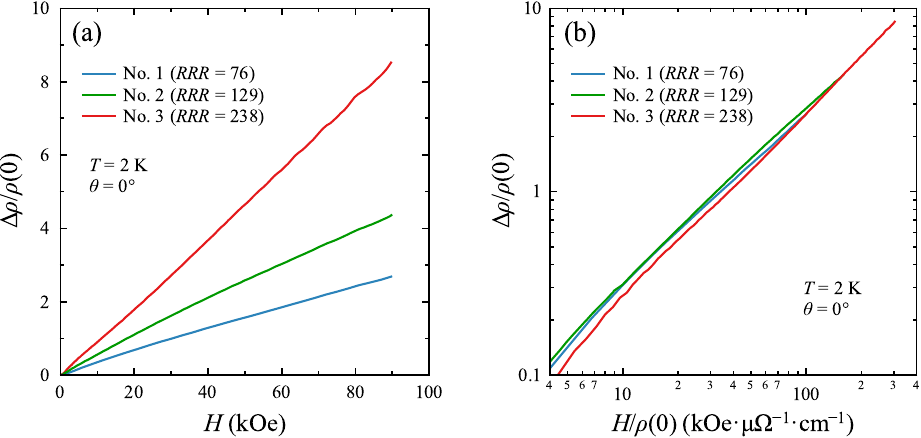}
        \caption{(a) $H$-dependence of transverse magnetoresistance  $\Delta \rho/\rho(0)$ measured at 2~K for different samples.
            (b) Kohler's plot of different \irsn\ samples.}
        \label{fig:sample-dep-MR}
    \end{figure}

    Figure~\ref{fig:sample-dep-MR}(a) shows $H$-dependence of transverse magnetoresistance (MR), defined by $\Delta \rho / \rho(0) \equiv [\rho(H)-\rho(0)]/\rho(0)$, 
    measured at 2~K for different samples with residual resistivity ratio ($RRR$) ranging from 76 to 238; sample No.~3 ($RRR=238$) was selected for the electronic transport measurements as reported in the main text.
    The linear-in-\textit{H} behavior is observed for all the samples (for $I\parallel[110]$ and $H\parallel[001]$), demonstrating that LMR is an intrinsic property of \irsn.
    Notably, samples with high $RRR$ exhibit Shubnikov--de Haas oscillations in high fields.
    
    Figure~\ref{fig:sample-dep-MR}(b) shows a $\Delta \rho / \rho(0)$ vs $H/\rho(0)$ plot (a Kohler's plot) for samples with different \textit{RRR}.
    The almost linear curves demonstrate that LMR occurs in a wide range of magnetic fields, spanning nearly two orders of magnitude.

    \clearpage
    \section{Carrier Mobility}
\begin{figure}[h]
        \centering
        \includegraphics{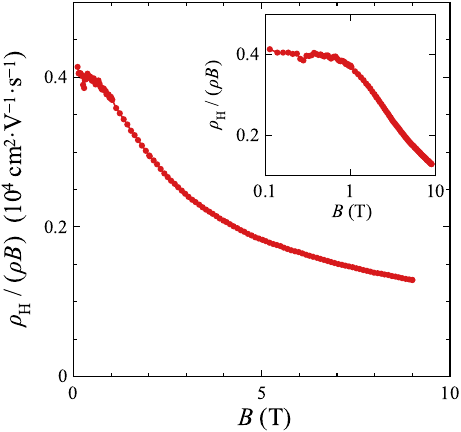}
        \caption{$B$-dependence of $\rho_\text{H}/(\rho B)$ measured at 2~K with $\theta=0^\circ$ for sample No. 3.}
        \label{fig:Hall-angle}
    \end{figure}

    \irsn\ is a compensated metal, which has equal numbers of electrons and holes.
    Therefore, the two-carrier model can be used for a reasonable estimation of the mobility.
    In this model, resistivity $\rho$ and Hall resistivity $\rho_\text{H}$ can be expressed as 
    \begin{equation}\label{eq:rho}
        \rho\equiv\rho_{xx} = \frac{n_\text{e}\mu_\text{e} + n_\text{h}\mu_\text{h} + (n_\text{e}\mu_\text{h} + n_\text{h}\mu_\text{e})\mu_\text{e}\mu_\text{h} B^2}{e(n_\text{e}\mu_\text{e} + n_\text{h}\mu_\text{h})^2 + e(n_\text{h} - n_\text{e})^2\mu_\text{e}^2\mu_\text{h}^2 B^2}
    \end{equation}
    and
    \begin{equation}\label{eq:rhoh}
        \rho_\text{H}\equiv\rho_{yx} = \frac{(n_\text{h}\mu_\text{h}^2 - n_\text{e}\mu_\text{e}^2) + (n_\text{h} - n_\text{e})\mu_\text{e}^2\mu_\text{h}^2 B^2}{e(n_\text{e}\mu_\text{e} + n_\text{h}\mu_\text{h})^2 + e(n_\text{h} - n_\text{e})^2\mu_\text{e}^2\mu_\text{h}^2 B^2}B,
    \end{equation}
    where $n_\text{e} (n_\text{h})$ and $\mu_\text{e} (\mu_\text{h})$ indicate the carrier concentrations and carrier mobilities of electrons (holes), respectively; $B=\mu_0 H$, where $\mu_0$ is the permeability of free space.
    However, as shown in Fig.~4(a) in the main text, $\rho$ exhibits LMR; and this cannot be reproduced by the abovementioned model equation with $n_\text{e} = n_\text{h}$, which exhibits a $B^2$ dependence.
    This fact indicates the necessity of some modification to this model.
    Nevertheless, we try to estimate the mobility %
    using Eqs. (\ref{eq:rho}) and (\ref{eq:rhoh}):
    \begin{equation}
        \frac{\rho_\text{H}}{\rho B} \left(= \frac{\tan\theta_\text{H}}{B}\right) \approx \mu_\text{h} - \mu_\text{e} \quad \text{when } B \rightarrow 0,
    \end{equation}
    where $\theta_\text{H}$ denotes the Hall angle. 
    As shown in Fig.~\ref{fig:Hall-angle}, %
    $\rho_\text{H}/(\rho B)$ saturates to $0.4 \times 10^4$~cm$^2$/(V$\cdot$s)
    below 1~T.
    This fact sets a lower bound on $\mu_\text{h}$, i.e., ${\mu_\text{h}>4000}$~cm$^2$/(V$\cdot$s); the high value of $\mu_\text{h}$ is in good agreement with high \textit{RRR} of the present crystal.

    \clearpage
    \section{Specific Heat}
\begin{figure}[h]
        \centering
        \includegraphics[width=.9\linewidth]{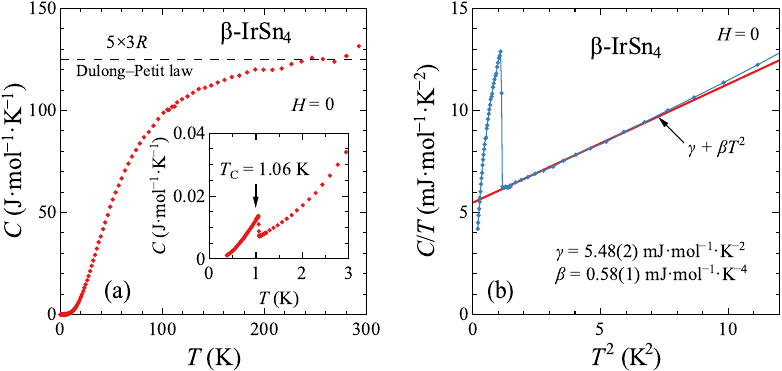}
        \caption{(a) Temperature-dependent specific heat $C$ of \irsn\ in zero field.
            The horizontal dashed black line represents the classical Dulong--Petit value of $C$. 
            The inset shows the SC jump appearing at $T_\text{C}=1.06$~K. 
            (b) The specific heat $C$ divided by the temperature $T$ as a function of $T^2$ at low temperatures. 
            The electronic specific heat coefficient $\gamma$ and the Debye temperature $\Theta_\text{D}$ can be evaluated from a linear fit of $\gamma + \beta T^2$ at low temperatures.}
        \label{fig:ct}
    \end{figure}

    Figure~\ref{fig:ct}(a) shows the $C$ vs. $T$ plot in zero field. 
    At 300 K, the value of $C$ is close to $3nR=124.65$, where $n=5$ (the number of atoms per formula unit) and $R = 8.31$~J/(mol$\cdot$K$^2$) (the gas constant), being in good agreement with the Dulong--Petit law.
    The inset shows the SC jump appearing at $T_\text{C}=1.06$ K. 
    Figure~\ref{fig:ct}(b) shows the specific heat $C$ over the temperature $T$ as a function of $T^2$ in low-temperature range, where a clear linear relation is observed below 2.5~K in the normal state. A linear fit of $\gamma + \beta T^2$ gives the Sommerfeld coefficient $\gamma=5.48$~mJ/(mol$\cdot$K$^2$) and the phononic coefficient $\beta=0.58$~mJ/(mol$\cdot$K$^4$).
    The Debye temperature $\Theta_\text{D}=256$~K is then estimated from $\Theta_\text{D} = [(12\pi^4/5\beta)nR]^{1/3}$.
    
    With $\Theta_\text{D}$ and $T_\text{C}$, the electron-phonon coupling constant $\lambda_\text{e-p}$ can be calculated by the inverted McMillan equation~\cite{mcmillan1968}:
    \begin{equation*}
        \lambda_\text{e-p}=\frac{1.04+\mu^*\ln (\Theta_\text{D}/1.45T_\text{C})}{(1-0.62\mu^*)\ln (\Theta_\text{D}/1.45T_\text{C})-1.04},
    \end{equation*}
    where $\mu^*=0.13$ is the empirical value of the Coulomb pseudopotential.
    It gives $\lambda_\text{e-p}=0.46$, suggesting weak-coupling SC, which is in line with the $\alpha$-model analysis mentioned in the main text.
